\documentclass[iop]{emulator}
\usepackage{amsmath}
\usepackage{natbib}
\newcommand{\Msun}{M$_{\odot}$ }

\slugcomment{To appear in ApJ Letters}

\shorttitle{The First Science Results from SPHERE: Disproving the Predicted Brown Dwarf around V471 Tau}
\shortauthors{Hardy et al.}

\begin{document}

\title{The First Science Results from SPHERE: Disproving the Predicted Brown Dwarf around V471 Tau\altaffilmark{*}}

\author{
A. Hardy\altaffilmark{1,6}, M.R. Schreiber\altaffilmark{1,6}, S.G. Parsons
\altaffilmark{1}, C. Caceres\altaffilmark{1,6}, 
G. Retamales\altaffilmark{1,6}
Z. Wahhaj\altaffilmark{2}, D. Mawet\altaffilmark{2}, 
H. Canovas\altaffilmark{1,6},
L. Cieza\altaffilmark{5,6},
T.R. Marsh\altaffilmark{3},
M.C.P. Bours\altaffilmark{3},
V.S. Dhillon,\altaffilmark{4},
A. Bayo\altaffilmark{1,6}
}

\altaffiltext{*}{Based on observations collected at the European Southern Observatory, Chile, program ID 60.A-9355(A)}

\altaffiltext{1}{Departamento de F\'{i}sica y Astronom\'{i}a, Universidad Valpara\'{i}so, Avenida Gran Breta\~{n}a 1111, Valpara\'{i}so, Chile.}

\altaffiltext{2}{European Southern Observatory, Casilla 19001, Vitacura, Santiago, Chile.}

\altaffiltext{3}{Department of Physics, Gibbet Hill Road, University of Warwick, Coventry, CV4 7AL, UK}

\altaffiltext{4}{Department of Physics and Astronomy, University of Sheffield, Sheffield S3 7RH, UK}

\altaffiltext{5}{Nucleo de Astronomia, Universidad Diego Portales, Av. Ej\'{e}rcito 441, Santiago, Chile.}

\altaffiltext{6}{Millennium Nucleus `Protoplanetary Disks in ALMA Early Science', Universidad de Valpara\'{i}so, Avenida Gran Breta\~{n}a 1111, Valpara\'{i}so, Chile}

\begin{abstract}
\noindent
Variations of eclipse arrival times have recently been detected in several post common envelope binaries consisting of a white dwarf and a main sequence companion star. The generally favoured explanation for these timing variations is the gravitational pull of one or more circumbinary substellar objects periodically moving the centre of mass of the host binary. 
Using the new extreme-AO instrument SPHERE, we image the prototype eclipsing post-common envelope binary \object{V471\, Tau} in search of the brown dwarf that is believed to be responsible for variations in its eclipse arrival times. We report that an unprecedented contrast of $\Delta m_{H}=12.1$ at a separation of 260\,mas was achieved, but resulted in a non-detection. This implies that there is no brown dwarf present in the system unless it is three magnitudes fainter than predicted by evolutionary track models, and provides damaging evidence against the circumbinary interpretation of eclipse timing variations. In the case of V471\,Tau, a more consistent explanation is offered with the Applegate mechanism, in which these variations are prescribed to changes in the quadrupole moment within the main-sequence star.
\end{abstract}

\keywords{stars: individual(V471 Tau) --- planet-star interactions --- binaries: eclipsing --- brown dwarfs --- white dwarfs --- binaries: close}

\section{Introduction}

Circumbinary substellar objects, although anticipated for a long time, have only recently been identified around main-sequence binary stars \citep{Doyle2011}. 
Long before this however, claims have been made for circumbinary substellar objects around close white dwarf-main sequence binaries. Initially consisting of a main sequence binary with separations of order $\sim$1 AU \citep{Zorotovic2013}, these systems are believed to have been through a spectacular phase of binary star evolution to explain their current close separation. When the more massive star of the binary evolves off the main sequence, it fills its Roche-lobe and generates dynamically unstable mass transfer onto the secondary star. As the time scale for dynamically unstable mass transfer is much shorter than the thermal time scale of the secondary, the latter cannot adjust its structure fast enough to incorporate the overflowing mass. Instead, a common envelope of material forms around both the secondary star and the core of the giant. 
Drag forces between the envelope and the central binary then cause the
envelope to be expelled at the expense of orbital energy and angular momentum 
of the binary \citep[e.g.][]{Webbink1984,Zorotovic2010, Ivanova2013}. 
The emerging white dwarf-main sequence binaries contain separations of just a
few solar radii, and are known as 
post common envelope binaries (PCEBs) \citep{Nebot2011}. 

Shortly after the discovery of the first PCEB 
it was realised that it displays variations in its eclipse arrival times. 
Today, similar variations are seen in almost all eclipsing PCEBs with long
enough coverage \citep{Parsons2010a, Zorotovic2013}, for which the most common
hypothesis is the presence of a circumbinary object, typically a brown dwarf
or multiple giant planets. In this 
scenario, the gravitational pull of the circumbinary objects periodically move
the center of mass of the host binary stars, thereby changing the light travel
time of the eclipse signal 
to Earth \citep{Irwin1959}. 
Indeed, the planetary model employed to explain the eclipse timing variations
(ETVs) seen in the PCEB NN Ser \citep{Beuermann2010} successfully predicted
new eclipse arrival times \citep{Beuermann2013, Marsh2014}, providing 
support to the circumbinary interpretation
but raising questions regarding the formation of these third objects. 
\citet{Zorotovic2013} favour a scenario in which the circumbinary objects form
as a consequence of the common envelope evolution, in a so-called second
generation scenario. This is based on the finding that nearly all PCEBs
with long enough coverage show ETVs, yet only a small fraction of main
sequence binaries seem to host 
circumbinary substellar objects. Indeed, \citet{Schleicher2014} were able to
develop a model in which a 
second generation protoplanetary disk forms during common envelope evolution
and produces giant planets through 
the disk instability model. 
In contrast, \citet{Bear2014} prefer the first generation scenario, in which
the objects form at a similar time to their main-sequence hosts, and survive
the common-envelope phase. They claim that if a second generation scenario
were true, too large a 
fraction of the common envelope mass would have to form into substellar
companions. 
%and this fraction is higher than is believed possible. %Also menion possible stable progenitor systems? 

However, before investigating possible formation scenarios further, we must
exercise caution with the third body hypothesis. 
Although the circumbinary object model has proved successful in the case of
NN\,Ser, this is an exception. 
In general the predictions from proposed planetary systems around PCEBs
disagree with more recent eclipse timing measurements \citep{Parsons2010a,
  Bours2014}, and some of the proposed planetary systems are dynamically
unstable 
on very short time scales \citep{Wittenmyer2013, Horner2012, Horner2013}. The
failure of all circumbinary object 
models except the one for NN Ser implies either that our timing coverage is
insufficient, or that there must 
be an alternative mechanism driving ETVs.

To progress with this situation, 
it has become vital that the circumbinary companion interpretation be tested
independently. The most conclusive 
way to achieve this is to image one of the proposed objects and the natural
choice for such an observation 
is V471\,Tau. V471\,Tau consists of a 0.84$\pm$0.05\,\Msun white dwarf and a
0.93$\pm$0.07\,\Msun secondary star 
\citep{OBrien2001}, and is a member of the 625\,Myr old Hyades open cluster 
\citep{Perryman1998}. Soon after its discovery \citep{Nelson1970}, 
\citet{Lohsen1974} reported ETVs which have been interpreted as being caused
by a circumbinary brown dwarf \citep{Beavers1986, Guinan2001}. 
V471\,Tau is ideal to test the circumbinary interpretation because it is
nearby, bright, and the proposed 
brown dwarf reaches projected separations exceeding 200\,mas, making detection
possible with the 
new extreme-AO facilities such as SPHERE \citep{Beuzit2008}.  

Here we present new high-precision eclipse times of V471\,Tau, and use these
to refine the proposed brown dwarf parameters using the Markov Chain Monte
Carlo (MCMC) method. We then test the circumbinary interpretation of ETVs with
SPHERE science verification observations, with high enough contrast to detect
the brown dwarf independent 
of if it formed in a second or first generation scenario. 

\section{Observations and Data Reduction}

\subsection{High-speed eclipse photometry}

In order to refine the orbit of the circumbinary companion we obtained
high-speed photometry of the eclipse of 
the white dwarf in V471\, Tau with the frame-transfer camera ULTRACAM
\citep{Dhillon2007} mounted as a visitor 
instrument on the 3.6-m New Technology Telescope (NTT) on La Silla in November
and December 2010. 
ULTRACAM uses a triple beam setup allowing one to obtain data in the $u'$,
$g'$ and $i'$ bands simultaneously, 
with negligible dead time between frames. Due to the brightness of the target
we de-focussed the telescope and 
used exposures times of $\sim$5\,s. There were no suitably bright nearby stars
in ULTRACAM's field of view to use 
for comparison. We therefore used the $i'$ band observations, where the
eclipse is extremely shallow, as a comparison source for the $u'$ band
data. This results in a large colour term in the resulting light curve, but
does remove variations in the conditions and does not affect the eclipse
timings. In some observations the conditions were good enough that no
comparison source was required. 

These data were reduced using the ULTRACAM pipeline software
\citep{Dhillon2007} and the resultant eclipse light 
curves were fitted with a dedicated code designed to fit binaries containing
white dwarfs 
\citep{Copperwheat2010}. The measured eclipse times were then heliocentrically corrected 
and are listed in Table. 1.

\begin{deluxetable}{cccc}
\tabletypesize{\scriptsize}
\tablecaption{ULTRACAM eclipse times for V471\, Tau\label{tbl-1}}
\tablewidth{0pt}
\tablehead{
\colhead{Cycle} & \colhead{HMJD(TDB)} & \colhead{Uncertainty (seconds)}
}

\startdata
2848 &  5512.2840584 & 1.76 \\
2886  & 5532.0889885 & 1.59 \\
2911 & 5545.1185942 &  1.62  \\
2915 & 5547.2033608 &  2.37  \\

\enddata

\end{deluxetable}

\subsection{SPHERE observations}

The imaging data testing the existence of the brown dwarf were acquired using
the extreme adaptive optics instrument, SPHERE at the UT3 Nasmyth focus of the
VLT, on 2014 December 11. An earlier set of observations was
  performed on 2014 December 6, but is not used here because of poorer data quality.
SPHERE is able to provide H and K-band images with Strehl ratios $>90\%$. The integral field spectrograph (IFS) and infrared dual-band imager and spectrograph (IRDIS) were used simultaneously in the IRDIFS mode. The IFS delivered a datacube containing 38 monochromatic images at a spectral resolution of R$\sim$50 between 0.96-1.34\,$\mu$m, whilst IRDIS delivered dual-band imaging in the H2 and H3 bands (central wavelengths of 1.59\,$\mu$m and 1.67\,$\mu$m respectively, and FWHMs of 0.0531\,$\mu$m and 0.0556\,$\mu$m). 

The brightness of the target and desired contrast required the use of the
N\_ALC\_YJH coronagraph, with an inner working angle of 0.15''. Detector
integration times were set at 64\,s for each detector. The observations were
obtained in pupil-stabilized mode, where the  
field rotates. To derotate and combine the images, one needs to accurately 
measure the center of rotation which is also the location
of the star behind the coronagraph. To achieve this, a waffle pattern was introduced into the
deformable mirror 
of the AO system, placing four replicas of the star equidistant from it in a
square pattern. 
These calibration images were taken before and after
the science, allowing subpixel accuracy in centroiding. Off-coronagraph,
unsaturated images of the star were 
also obtained with a neutral density
filter to allow sensitivity/contrast measurements. Basic reduction, including
dark and flat-fielding was performed using the SPHERE pipeline. 
Custom angular differential imaging (ADI) routines, particularly for subpixel shifting and aligning of speckles,
were used to subtract the star \citep{Wahhaj2013}. 
A custom principal component analysis routine was also applied to the data 
to compare with the ADI reduction \citep{Mawet2013}.

\section{Predicting the projected separation of the potential brown dwarf}

Assuming a third body orbiting around V471\, Tau, the time delay or advance
caused by this body can be expressed as
%\begin{center}
\begin{equation} 
\Delta T=\frac{a_{12}\,\sin\,i_{3}}{c}\left [\frac{1-e^{2}_{3}}{1+e_{3}\,\cos\nu_{3}}\sin(\nu_{3}+\omega)+e_{3}\,\sin(\omega)\right ]
\end{equation} 
%\end{center}
\citep[e.g.][]{Irwin1959} where $c$ is the speed of light and $a_{12}$ is the
semi-major axis of the binary star's orbit around the common center of mass of
the triple system. The other parameters define the orbit of the third body,
i.e. its inclination $i_{3}$, the orbital eccentricity $e_{3}$, argument of
periastron $\omega$, and true anomaly $\nu_{3}$.  

As shown by \cite{Marsh2014}, strong correlations can exist between orbital
parameters and the problem is highly degenerate 
unless a large number of high precision eclipse timing measurements are
available. Only our recent ULTRACAM measurements provide precise eclipse timings, with uncertainties of $\sim$1.8\,s, whereas the timings in
  the literature have been assigned large error estimates of 15\,s for the sake
  of caution. To properly identify not only the
best fit parameters but also their uncertainties we performed a
Markov Chain Monte-Carlo (MCMC) simulation to the eclipse times. 

The prediction of the best-fit model can be seen in Fig.\,1, (top left panel)
alongside all archival observed-minus-calculated eclipse times
  \citep{Kundra2011} and the new times reported in this paper. 
This best-fit model corresponds to a brown 
dwarf of mass 0.044$\pm$0.001 \Msun and semi-major axis 12.8$\pm$0.16 AU. 

\begin{figure*}
\epsscale{0.90}
\plotone{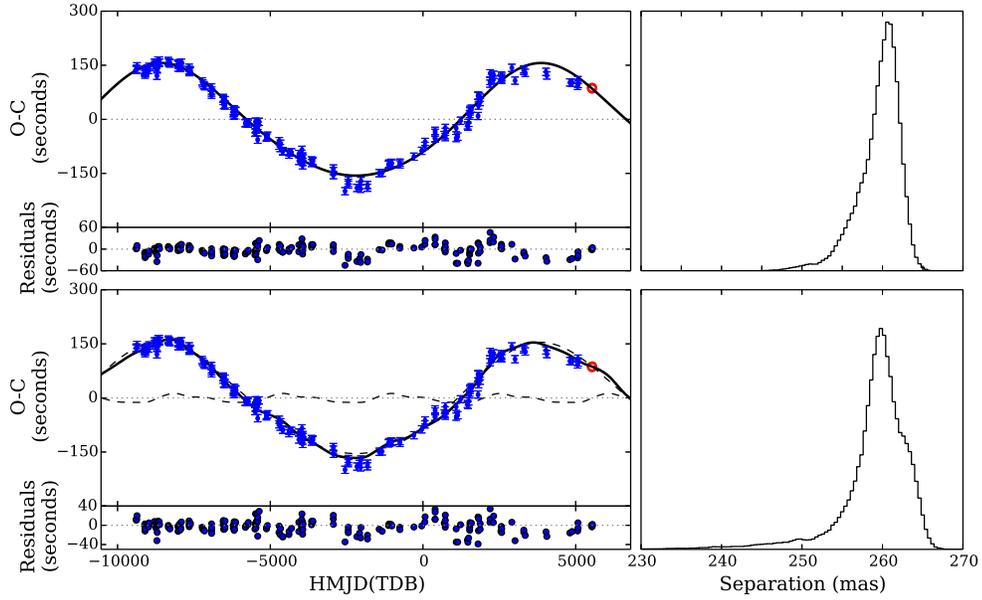}
\caption{Observed-minus-calculated (O-C) eclipse arrival times for V471\,
  Tau. Upper left panel: The eclipse times and associated errors, with
    the high precision eclipse times reported in this paper as the red open
    circles. The best-fit model assuming a third-body perturbation shown as
  the solid black line. The residuals of this fit can be seen directly beneath
  the curve.  
Upper right panel: The projected separations predicted by the MCMC simulation
for observation in December 2014, assuming a distance of 50 pc to V471
Tau. The lower panels are identical, but denote the results 
of a 2 body fit the the eclipse timing variations. The dashed lines in the
lower left panel denote the contributions from the different objects, with
their sum in black. Despite the extra body, the predicted separation and mass of the
brown dwarf hardly change. \label{fig1}} 
\end{figure*}

While the parameters of the brown dwarf in the one body fit are well
constrained by the eclipse times (Fig.\,2), 
the residuals are far from random and suggest another mechanism may also be at work.   
To test this possibility, we performed another MCMC with 2 companions to
account for these deviations. This further
allowed us to test whether the brown dwarf causing the main variation could be
at a smaller separation 
or be less massive, which would make it harder to detect. 
The resulting best fit is shown in the bottom left panel of Fig.\,\ref{fig1}.
 
\begin{figure*}
\epsscale{.60}
\plotone{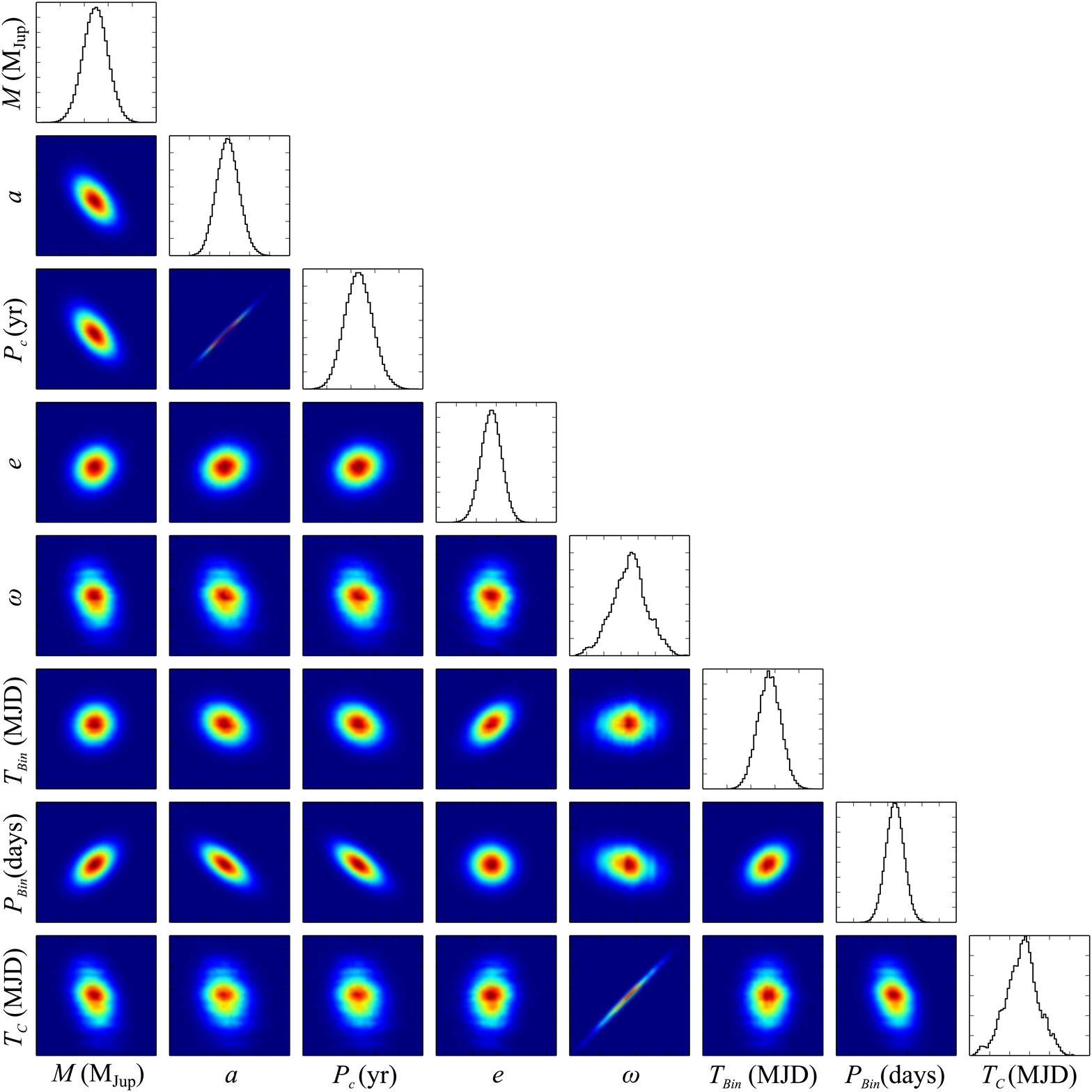}
\caption{Correlations and histograms for the Markov-chain Monte Carlo simulation. \textit{M} is the mass of the brown dwarf, \textit{a} the semi-major axis of its orbit, \textit{P$_{C}$} the period, \textit{e} the eccentricity, \textit{$\omega$} the argument of periastron, \textit{T$_{Bin}$} the time of periastron of the binary, \textit{P$_{Bin}$} the period of the binary, and \textit{T$_{C}$} the time of periastron of the brown dwarf. All parameters appear well constrained with no significant correlations. \label{fig2}}
\end{figure*}

The derived orbital parameters for both cases were then projected onto the sky
using a distance to V471\,Tau of $\sim$50 pc, as measured by Hipparcos
\citep{vanLeeuwen2007}, to obtain a predicted separation for the brown dwarf 
companion in December 2014.  The simulation suggests a separation of
$260\substack{+6 \\ -19}$\,mas for the one body fit and this value hardly
changes if an additional object is assumed to account for the problems of 
the one body fit (see Fig. 1, right panels).

\section{Testing the prediction with SPHERE}

No third component is present in the SPHERE IRDIS images (Fig. 3, left
panel). The contrast achieved was estimated via two different methods of fake
  companions injection. In the first, fake companions of a known contrast were
  injected at different angular separations, and the contrast defined by where
  the fake companion was recovered 95\% of the time \citep{Wahhaj2013}. In the
  second method, the fake companions were used to measure the post-ADI
  throughput loss, and this was used to renormalize the contrast curve of 
a typical saturated SPHERE PSF. In both cases, the H2 and H3 channels were
summed as no spectral difference 
between the channels was expected, and the curves were 
corrected for small-sample statistics at small separations \citep{Mawet2014}. 
The resulting contrast curves for the IRDIS detector can be seen
in Fig. 3, right panel, with the former method as the solid line, and the latter as the dashed. There is good agreement between the 2 methods at the
predicted separation of $\sim$260\,mas, and both indicate an achieved contrast
of $\sim$12.1 magnitudes. To determine if this is sufficient to indeed detect
the brown dwarf, it is necessary to know its age, mass and metallicity, from
which the brown dwarf luminosity can be predicted. We find the mass is well
constrained from the MCMC models, and the metallicity was assumed identical to
other members of the Hyades cluster, with [M/H]=0.14$\pm$0.05
\citep{Perryman1998}.

\begin{figure*}

\epsscale{.80}
\plotone{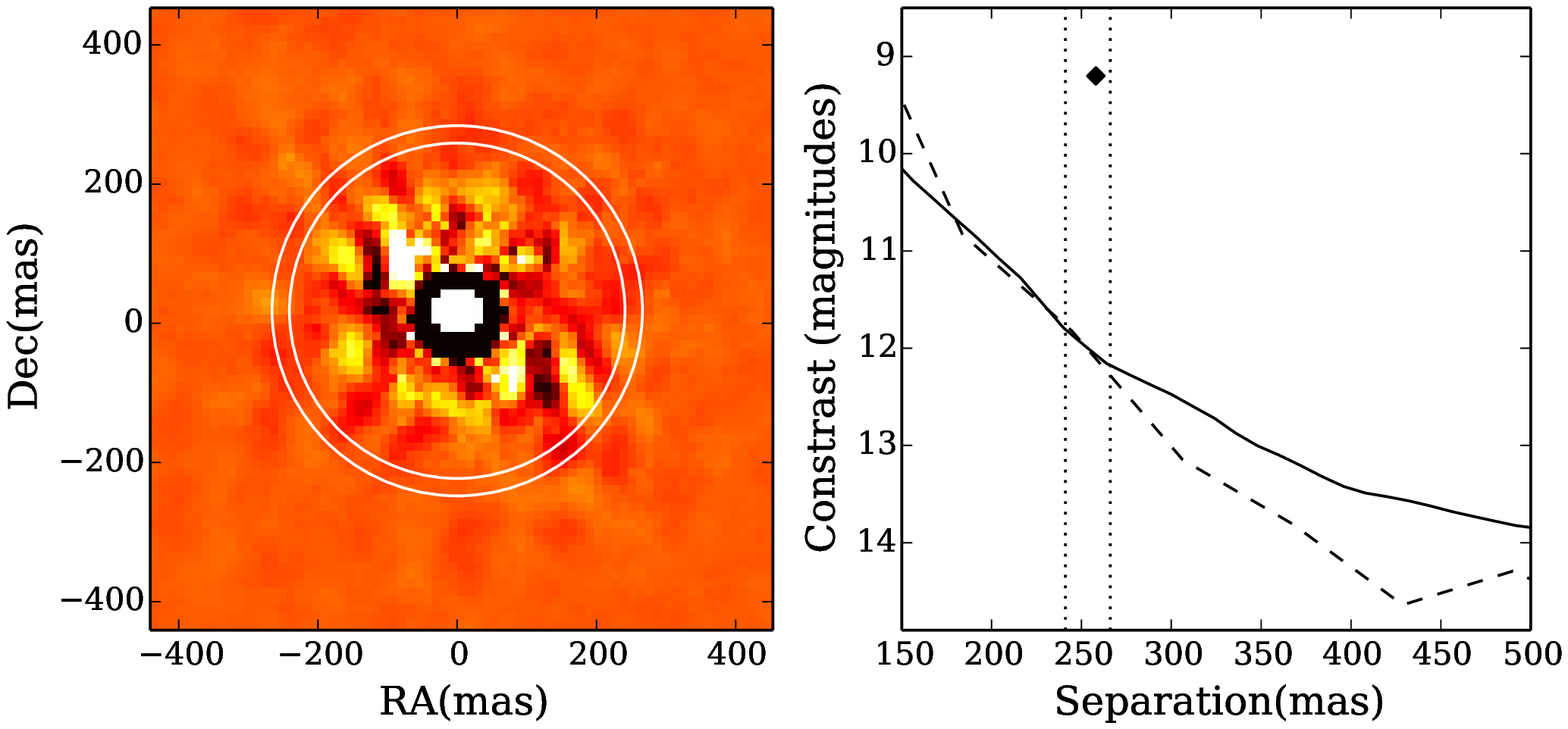}
\caption{H-band image of V471\, Tau obtained on the SPHERE IRDIS instrument at
  the VLT. Left panel: Resulting image after angular differential imaging
  (ADI). The area in-between the white circles denotes the 5 sigma predicted
  position of the brown dwarf. Right panel: Contrast curves obtained
    via 2 different methods of fake companion injection (see text for
    details). The vertical lines again denote the predicted position of the
  object, with 
the diamond denoting the predicted contrast of a first generation brown dwarf.\label{fig3}}

\end{figure*}

The cooling age of the white dwarf in V471\,Tau is $\sim10$\,Myrs, and puts a
stringent constraint on the age in a 
second generation scenario. If the 
$0.044$\,\Msun brown dwarf had formed in such a scenario, BTSettl models
\citep{Allard2012} combined with isocrones from \citet{Baraffe1998} predict a $\Delta m_{H}\sim$ 4.5. This is 7.5 magnitudes brighter than 
our detection limit, conclusively ruling out a second generation formation
scenario for the potential brown 
dwarf around V471\,Tau.  

If the brown dwarf formed in a first generation scenario, we can obtain an
estimate of its age from the age of the Hyades cluster (625\,Myr). An identical
modelling procedure suggests that such a brown dwarf will have a contrast
of $\Delta m_{H}\sim$ 9.2 in the H band, 3 magnitudes brighter than our
detection limit.

\section{Discussion}

\subsection{Could the BD have escaped our SPHERE observations?} 

Inspecting the contrast curve presented in Fig.\,3, it is clear that 
if a first generation brown dwarf was at a smaller separation and/or had a
lower mass, it may still have escaped detection. To
test whether a brown dwarf could escape detection whilst simultaneously  causing the O-C variations, we performed a final MCMC simulation with separations limited to only those which would be undetectable by SPHERE. 
The resulting fit drastically failed to explain the O-C data, and delivered a $\chi^2_{red}$ of 47.7,
compared to 1.1 in the case of the constraint-free brown dwarf. As shown in Fig.\,1, (lower left panel), not
even the addition of an extra body can cause the brown dwarf to reach a
separation undetectable by SPHERE.   
In order to recreate the observed O-C variations, a large mass, high
separation companion seems needed, and we therefore find no configuration in which the brown dwarf would have escaped SPHERE. 

The second possibility as to how the brown dwarf might have escaped detection,
is that the models of substellar evolutionary tracks are incorrect. Indeed, if
the 0.044\,\Msun brown dwarf was at a temperature of $\sim$700\,K, as oppose to
the $\sim$1500\,K predicted for a first-generation brown dwarf, it would escape
detection. However, despite the uncertainties that are associated with these
cooling models, there is no evidence to suggest 
models are off to this degree \citep{Bell2012, Lodieu2014}. 

Finally, the brown dwarf around V471\,Tau might be unique because of the
evolutionary history of the host binary star. One might for example speculate
that it accreted significant amounts of material during the rather recent
common envelope phase, and this could have caused its metallicity to deviate
significantly from the metallicity of the Hyades. Indeed, if the brown dwarf
were first generation, and possessed a metallicity [M/H] = 0.5, models predict
it would not be detected in the presented observations
\citep{Allard2011}. However, the metallicity of the secondary K star has been
studied, and found to show no peculiar abundances despite the possibility it accreted material during the common envelope phase
\citep{Still2003}. It is therefore highly unlikely that the brown dwarf
metallicity is high enough to explain the non-detection. A final effect of
recent accretion, could be that the infalling mass made the brown dwarf
fainter due to contraction \citep{Baraffe2009}, although confirming if this is indeed possible will require detailed evolutionary brown dwarf models dedicated to V471\,Tau  which is beyond the scope of the present paper.

\subsection{Applegate's mechanism}

If our non-detection is due to the lack of a brown dwarf, an alternative
mechanism must be responsible for the ETVs. Currently, there are two such
alternative theories for ETVs in PCEBS. The first, apsidal precession, can
confidently be ruled out for 
V471\,Tau. This mechanism prescribes ETVs to the time dependance in the argument of periastron. Although apsidal precession would require an eccentricity in the binary of just e=0.01 to create the $\sim$300\,s timing variations seen in Fig.\,\ref{fig1}, the timescale would be less than 3 years, and not $\sim$30 years as observed. The second alternative explanation, the Applegate mechanism \citep{Applegate1992}, is potentially able to drive the variations
of the eclipse arrival times seen in V471\,Tau. This theory prescribes these
variations to quasi-periodic oblateness changes in the main sequence star,
presumed to be driven by solar-like magnetic cycles. The K star in V471\,Tau
is particularly active, and may provide sufficient energy to drive these
variations. Applegate's mechanism is therefore the currently most convincing
explanation for the eclipse arrival times observed in V471\,Tau.

However, although a suitable explanation in the case of V471\,Tau, in almost
all other PCEBs showing ETVs, 
classical Applegate's mechanism can be ruled out
as they tend to contain less active 
main-sequence stars. NN Ser is one such system, and currently only the proposed planetary system can explain its behavior \citep{Parsons2014, Brinkworth2006, Parsons2010}, although it is possible that an as-yet unconsidered model of magnetic field variations could still offer an explanation.
Thus, Applegate does not offer a comprehensive explanation for ETVs, and
confirmation of its effect 
in V471\,Tau is needed alongside additional tests of the third body interpretation in other systems.

\section{Conclusion}

We have presented deep SPHERE science verification observations of 
V471\,Tau testing the hypothesis that the observed ETVs are caused by a
circumbinary brown dwarf. We reached an excellent contrast  
of $\Delta m_{H}=12.1$ at the predicted separation of the brown dwarf but no companion can be
seen in the images. This excludes both a brown dwarf formed in a second
generation scenario, 
as well as a standard brown dwarf at the age of the Hyades cluster to be
present around V471\,Tau. The Applegate 
mechanism is hence the one and only remaining model currently explaining the ETVs seen in V471\,Tau. 

With this result, the origin of ETVs in PCEBs remains a puzzle. While no theory but the existence of two circumbinary planets can 
currently explain the variations seen in the PCEB NN\,Ser, the most reasonable
explanation for the variations seen in V471\,Tau is now the Applegate
mechanism. We therefore conclude that in their current form neither 
the third body interpetation nor Applegate's mechanism offer a general
explanation for the ETVs observed in 
nearly all PCEBs.

\acknowledgments

AH, MRS, CC, HC, LC and AB acknowledge support from the Millennium Nucleus RC130007 (Chilean Ministry of Economy). MRS, SP, CC, LC and AB also acknowledge support from FONDECYT grants 1141269, 3140585, 3140592, 1440109 and 11140572 respectively. TRM thanks the UK's Science and Technology Facilities Council for support during the course of this work under grant ST/L000733/1.

{\it Facilities:} \facility{VLT (SPHERE)}, \facility{NTT (ULTRACAM)}.

\appendix

\end{document}